\documentclass[a4paper,11pt]{article}
\usepackage{latexsym}
\usepackage{epsfig, subfigure, subfloat, graphicx, float }
\usepackage{amssymb}
\usepackage{epstopdf}
\author{H. Mohseni Sadjadi\footnote{mohsenisad@ut.ac.ir}
\\ {\small Department of Physics, University of Tehran,}
\\ {\small P. O. B. 14395-547, Tehran 14399-55961, Iran}}
\author{H. Mohseni Sadjadi \footnote{mohsenisad@ut.ac.ir}
\\ {\small Department of Physics, University of Tehran,}
\\ {\small P. O. B. 14395-547, Tehran 14399-55961, Iran}}
\title{Generalized Noether symmetry in $f(T)$ gravity}
\begin{document}
\maketitle
\begin{abstract}
We consider modified teleparallel gravity ($f(T)$ gravity), as a
framework to explain the present accelerated expansion of the
universe. The matter component is assumed to be cold dark matter.
To find the explicit form of the function $f$, we utilize
{\it{generalized}} Noether theorem and use {\it{generalized}}
vector fields as variational symmetries of the corresponding
Lagrangian. We study the cosmological consequences of the obtained
results.
\end{abstract}

\section{Introduction}

In teleparallel gravity \cite{tel}, the gravitational interaction is
described using torsion, instead of the curvature used in general
relativity; and instead of the torsion-less Levi-Civita
connection, curvature-less Weitzenb\"{o}ck \cite{connect} connection is employed.
The gravitational action of this model is given by
\begin{equation}\label{1}
S_T={1\over 16\pi}\int \left|e\right|T d^4x,
\end{equation}
where $\left|e\right|=\it{det}\left(e^a_\mu\right)$, and the
metric components are related to tetrad via
\begin{equation}\label{2}
g_{\mu \nu}=\eta_{ab}e^a_\mu e^b_{\nu}.
\end{equation}
In (\ref{1}), the torsion scalar $T$ is
\begin{equation}\label{3}
T={1\over 2}\left({K^{\mu
\nu}}_{\sigma}+\delta^{\mu}_{\sigma}{T^{\alpha
\nu}}_{\alpha}-\delta^\nu_\sigma {T^{\alpha
\mu}}_\alpha\right){T^{\sigma}}_{\mu \nu},
\end{equation}
where
\begin{equation}\label{4}
{T^\sigma}_{\mu \nu}=e^{\sigma}_a\left(\partial_\nu
e_\mu^a-\partial_\mu e_\nu^a\right),
\end{equation}
is torsion of the Weitzenb\"{o}ck connection, and
\begin{equation}\label{5}
{K^{\mu \nu}}_\sigma={1\over 2}\left({T^{\nu
\mu}}_\sigma+{T_\sigma}^{\mu \nu}-{T^{\mu \nu}}_{\sigma} \right),
\end{equation}
is the contorsion tensor.

A candidate to describe the present accelerated expansion of our
universe\cite{acc}, is the modified theory of gravity
\cite{modif}. In modified theories based on general relativity,
the gravitational action $S={1\over 16\pi}\int R\sqrt{-g}d^4x$ is
replaced by $S={1\over 16\pi}\int f(R)\sqrt{-g}d^4x$. Inspired by
this model modified teleparallel gravity has been proposed to
study the acceleration expansion of our universe \cite{teled}.
This model is described
\begin{equation}\label{6}
S={1\over 16\pi}\int d^4x \left|e\right|f(T)+S_m,
\end{equation}
where $S_m$ is matter action.

In some papers, the form of $f(T)$ is suggested and then its
cosmological consequences are investigated \cite{lin}. In some
other papers, the Hubble parameter or the behavior of the
effective energy density is specified and then using modified
Friedmann equations, the form of $f(T)$ is obtained \cite{mir}.

Another way to determine a specific form for $f(T)$, is to use
symmetries of the problem. Noether symmetry provides us a mean to
get some insights about the form of $f(T)$ \cite{hao}. This method
was used to determine the form of the potential in quintessence
model, and to specify the modifications in modified theories of
gravity \cite{quinmod}. In the aforementioned papers, the studies
were restricted to Noether symmetries corresponding to vector
fields whose coefficients were assumed to depend only on time and
coordinates (in configuration space). In generalized symmetries,
the coefficient functions of the vector fields contain first and
higher order time derivatives of coordinates. In this situation, a
generalization of Noether theorem can be realized. To see a
discussion about this case see \cite{GN}, where the generalized
Noether symmetry and their corresponding generalized vector fields
are studied. To have an insight of this subject, consider the
motion of a particle under influence of a central force,
characterized by the Lagrangian
\begin{equation}\label{7}
L={1\over 2}\left[\left(\dot{r}^2+r^2\dot{\theta^2}\right)+{k\over
r}\right].
\end{equation}
In this problem, one can find the generalized vector fields \cite{kep}
\begin{eqnarray}\label{8}
X_1&=&r^2\cos\theta\dot{\theta}{\partial\over \partial
r}+\left(\cos\theta\dot{r}-2r\sin\theta\dot{\theta}\right){\partial\over
\partial \theta}\nonumber \\
X_2&=&r^2\sin\theta\dot{\theta}{\partial\over \partial
r}+\left(\sin\theta\dot{r}+2r\cos\theta\dot{\theta}\right){\partial\over
\partial \theta},
\end{eqnarray}
as variational symmetries of lagrangian related to Runge-Lenz
vector \cite{GN}.

In this manuscript, we consider modified teleparallel gravity
($f(T)$ model), as a framework to explain the present accelerated
expansion of the universe and intend to use generalized Noether
symmetry corresponding to generalized vector fields to find the
explicit form of the function $f(T)$.

The structure of the manuscript is as follows: In the second
section after some preliminaries, we introduce the Lagrangian
formalism for modified teleparallel gravity. We assume that the
dominant matter component is cold dark matter. Based on
generalized Noether theorem, we introduce a generalized vector
field and obtain a system of partial differential equations for
the coefficients and the function $f$. In the third section, the
system of differential equations is solved and based on explicit
form derived for $f(T)$, and integrals of motion, some
cosmological consequences of the obtained results are discussed.

We use units $\hbar=c=G=1$ through the paper.

\section{Generalized Noether symmetry}
\subsection{Preliminaries}
We consider a spatially flat Friedmann-Robertson-Walker (FRW)
space time in comoving coordinates
\begin{equation}\label{9}
ds^2=dt^2-a^2(t)(dx^2+dy^2+dz^2).
\end{equation}
$a$ is the scale factor and in terms of the Hubble parameter,
$H={\dot{a}\over a}$, the scalar torsion (\ref{3}) is given by
\begin{equation}\label{10}
T=-6H^2.
\end{equation}
The modified Friedmann equation is
\begin{equation}\label{11}
H^2={8\pi\over 3}\left({{\rho_m-{f\over 16\pi}}\over
2f_{,T}}\right)={8\pi\over 3}(\rho_m+\rho_T),
\end{equation}
where $f_{,T}={df\over dT}$, and the effective dark energy density
is
\begin{equation}\label{12}
\rho_T=-{1\over 16\pi}(T+f)+{Tf_{,T}\over 8\pi}.
\end{equation}
The Raychaudhury equation is given by
\begin{equation}\label{13}
48H^2f_{,TT}\dot{H}-f_{,T}(4\dot{H}+12H^2)-f=16\pi P_m,
\end{equation}
where $P_m$ is matter pressure. $\rho_m$ and $P_m$ satisfy The
continuity equation
\begin{equation}\label{14}
\rho_m+3H(P_m+\rho_m)=0.
\end{equation}
The equation of state parameter (EoS) of the universe can be
written as
\begin{eqnarray}\label{15}
w&=&-1-{2\over 3}{\dot{H}\over H^2}\nonumber \\
&=&-1+{2Tf_{,T}-f\over T(2Tf_{,TT}+f_{,T})}.
\end{eqnarray}
We assume that the matter component is dominated by cold dark
matter characterized by $P_m=0$, leading to
$\rho_{m}=\rho_{m0}a^{-3}$, where $\rho_{m0}$ is a constant.  By
adopting a suitable Lagrangian \cite{hao}
\begin{equation}\label{16}
\mathcal{L}=a^3(t)(f-f_{,T}T)-6f_{,T}a(t)\dot{a}^2(t)-16\pi\rho_{m0},
\end{equation}
in the configuration space ${q_i}=\{a,T\}$, the Euler Lagrange
equation ${d\over dt}{\partial \mathcal{L}\over
\partial \dot{T}}={\partial \mathcal{L}\over \partial T}$ results in $T=-6H^2$, when
\begin{equation}\label{17}
f_{,T}T\neq 0.
\end{equation}
The modified Raychoudhury equation (\ref{13}) is deduced from
${d\over dt}{\partial \mathcal{L}\over
\partial \dot{a}}={\partial \mathcal{L}\over \partial a}$.
The modified Friedmann equation is emerged from the Hamiltonian
constraint \cite{hao}. Indeed
\begin{equation}\label{18}
\sum_i {\partial \mathcal{L}\over \partial
\dot{q}_i}\dot{q}_i-\mathcal{L}=0,
\end{equation}
gives
\begin{equation}\label{1000}
12H^2f_{,T}+f={16\pi \rho_{m0}\over a^3},
\end{equation}
which is the same as (\ref{11}) rewritten for cold dark matter.
\subsection{Generalized symmetry}
A generalized vector field is expressed as \cite{GN,kep}
\begin{equation}\label{19}
X=\epsilon (t,q^i,\dot{q}^i,...){\partial \over \partial
t}+\sum_{j=1}^n \varphi^j (t,q^i,\dot{q}^i,...) {\partial \over
\partial q^j},
\end{equation}
where $\epsilon$ and $\varphi^j$ are smooth functions of $t$, $n$
canonical coordinates $q^i$, and their first and higher order time
derivatives. $X$ is the generator of a variational symmetry of the
Lagrangian iff there there exists a continuous function $B$ such
that
\begin{equation}\label{20}
pr^1X(\mathcal{L})+\mathcal{L}{d\epsilon    \over dt}={dB\over
dt},
\end{equation}
where the first prolongation of $X$ is
\begin{equation}\label{21}
pr^1X=X+\sum_{j=1}^n\left(
\dot{\varphi^j}-\dot{q^j}\dot{\epsilon}\right) {\partial \over
\partial q^j}.
\end{equation}
Inspired by (\ref{8}), and for the sake of simplicity, in the same
manner as \cite{GN}, we restrict ourselves to the case where the
coefficients are linear in the velocities
\begin{eqnarray}\label{22}
X&=&\epsilon(a,T){\partial\over \partial
t}+\left(\epsilon_1(a,T)+\alpha(a,T)\dot{a}+\beta(a,T)\dot{T}\right){\partial\over
\partial a}\nonumber \\
&&+\left(\epsilon_2(a,T)+\lambda(a,T)\dot{a}+\gamma(a,T)\dot{T}\right){\partial\over
\partial T}.
\end{eqnarray}
The Noether integral is
\begin{eqnarray}\label{23}
\mathcal{P}&=&B-\epsilon\mathcal{L}-\left(\epsilon_1+\alpha\dot{a}+\beta
\dot{T}\right){\partial \mathcal{L}\over \partial \dot{a}}-
\left(\epsilon_2+\lambda\dot{a}+\gamma \dot{T}\right){\partial
\mathcal{L}\over \partial \dot{T}}\nonumber
\\
&&+\epsilon\dot{a}{\partial \mathcal{L}\over
\partial \dot{a}}+\epsilon\dot{T}{\partial \mathcal{L}\over
\partial \dot{T}}.
\end{eqnarray}
By putting (\ref{16}) and (\ref{22}) in (\ref{20}), and equating
expressions containing the same order of time derivatives of $a$
and $T$ in both sides, we get:
\begin{equation}\label{24}
B=p(a,T)-6af_{,T}\alpha \dot{a}^2,
\end{equation}
\begin{equation}\label{25}
\beta=0,
\end{equation}
where $p(a,T)$ is a continuous function, and also a system of
differential equations:
\begin{eqnarray}\label{26}
&&\lambda f_{,TT}+\alpha_{,a}f_{,T}-\epsilon_{,a}f_{,T}=0,\nonumber \\
&&3a^2\alpha\left(f_{,T}T-f\right)+Ta^3\lambda f_{,TT}+\epsilon_{,a}a^3\left(f_{,T}T-f\right)+\rho_{m0}\epsilon_{,a}+p_{,a}=0,\nonumber \\
&&\gamma f_{,TT}-f_{,T}\epsilon_{,T}+f_{,T}\alpha_{,T}-\alpha f_{,TT}=0,\nonumber \\
&&Ta^3\gamma f_{,TT}+a^3\epsilon_{,T}\left(Tf_{,T}-f\right)+\rho_{m0}\epsilon_{,T}+p_{,T}=0,\nonumber \\
&&\epsilon_1f_{,T}+a\epsilon_2f_{,TT}+2af_{,T} \epsilon_{1,a}=0,\nonumber \\
&&3a^2\left(f_{T}T-f\right)\epsilon_1+a^3Tf_{,TT}\epsilon_2=0,\nonumber \\
&&12af_{,T}\epsilon_{1,T}=0.
\end{eqnarray}
The integral of motion is
\begin{equation}\label{27}
\mathcal{P}=p-12\epsilon_1a\dot{a}f_{,T}+a^3\epsilon\left(f_{,T}T-f\right)
-6a\dot{a}^2\left(\alpha-\epsilon\right)f_{,T}+16\pi \epsilon
\rho_{m0}.
\end{equation}

\section{Solutions}
$\alpha$, $\beta$, $\lambda$ and $\gamma$ do not involve in the
three last equations in (\ref{26}). These three equations are
sufficient to determine the form of $f(T)$ if either of
$\epsilon_1$ and $\epsilon_2$ is non zero. In this situation one
obtains a power law expression for $f(T)=\mu T^n$, which using
(\ref{1000}) leads to $a(t)\propto t^{2n\over 3}$ whose the
cosmological consequences are discussed in \cite{hao}. In this
case by solving (\ref{26}), an additional Noether symmetry
corresponding to the the generalized vector field specified by
\begin{equation}\label{28}
\alpha=F(y),\,\,\, \lambda=-{3T^2F'(y)a^{3-n\over n}\over
n(n-1)},\,\,\,\,\gamma=F(y)-{Ta^{3\over n}F'(y)\over n-1},
\end{equation}
where $y=Ta^{3\over n}$ and $F$ is an arbitrary continuous function, is attained.

For $\epsilon_1=\epsilon_2=0$, obtaining an analytical solution
for the system (\ref{26}) is very complicated, if not impossible.
So to go further, one should examine specific cases. Here we
consider solutions characterized by $\epsilon=0$. By this
simplification the following specific solutions for $f(T)$ are
derived (using the Maple 13 PDEtools package) :
\begin{eqnarray}\label{29}
f(T)&=&C_1T+C_2 \nonumber \\
f(T)&=&C_1\sqrt{-T}+C_2,\nonumber \\
\end{eqnarray}
which are not acceptable because the first one is not consistent
with $f_{,TT}\neq 0$ used in our procedure, and the second one
when inserted in (\ref{1000}) gives $\rho_m={C_2\over 16\pi}$,
which does not describe cold dark matter. The third specific
solution is
\begin{eqnarray}\label{30}
f(T)&=&\pm\sqrt{2C_1T+2C_2}\nonumber \\
\gamma&=&0 \nonumber \\
p&=&\pm {a^6C_3\over \sqrt{2}}+C_4\nonumber \\
\alpha&=&{C_3a^3\over \sqrt{C_1T+C_2}}\nonumber\\
\lambda&=&{6a^2C_3\sqrt{C_1T+C_2}\over C_1},
\end{eqnarray}
with the condition
\begin{equation}\label{31}
C_1T+C_2\geq 0,
\end{equation}
implying that $H$ is real.  For this solution, the Noether
integral is
\begin{equation}\label{32}
\mathcal{P}=p+6a\dot{a}^2\alpha f_{,T},
\end{equation}
which can be rewritten as
\begin{equation}\label{33}
\pm{C_2\over C_2-6C_1H^2}={d\over a^6},
\end{equation}
where the constant $d$ is defined by
$d:={(\mathcal{P}-C_4)\sqrt{2}\over C_3}$.

By using (\ref{1000}) and after some computations we obtain $d=\pm
{\tilde{\rho}_{m0}^2\over 2C_2}$, where $\tilde{\rho}_{m0}=16\pi
\rho_{m0}$, leading to
\begin{equation}\label{34}
{c_2\over c_2-6c_1h^2}={1\over 2c_2a^6}.
\end{equation}
Dimensionless parameters $c_1$ and $c_2$ are defined through
$C_2=c_2\tilde{\rho}_{m0}$; $C_1=\tilde{\rho}_{m0}c_1$. We use
$\tau=t\sqrt{\tilde{\rho}_{m0}}$ instead of the cosmic time and
dimensionless Hubble parameter $h={1\over a(\tau)}{da(\tau)\over
d\tau}$ is considered. (\ref{31}) may be rewritten as
\begin{equation}\label{35}
c_2-6c_1h^2\geq 0,
\end{equation}
and (\ref{34}) implies
\begin{equation}\label{36}
{c_2\over 2c_1}-{c_2^2\over c_1}a^6\geq 0.
\end{equation}

To see whether the phantom divide line ($w=-1$) crossing is
allowed in this model, we must compute ${dh\over d\tau}$. Using
(\ref{34}), after some calculations, we obtain
\begin{equation}\label{37}
{dh\over d\tau}=-{c_2^2a^6\over c_1}.
\end{equation}
Therefore the transition from ${dh\over d\tau}<0$ to ${dh\over
d\tau}>0$ and vice versa are not possible. In the following we
consider ${dh\over d\tau}<0$ which corresponds to $c_1>0$.
(\ref{35}) leads to $c_2>0$.  To study the cosmological
consequences of this model we try to solve the differential
equation (\ref{34}), with the constraint $\{c_1>0,c_2>0\}$. We
write (\ref{34}) as
\begin{equation}\label{38}
\left({da(\tau)\over d\tau}\right)^2={1\over 6}\left({c_2\over
c_1}-2{c_2^2a^6(\tau)\over c_1}\right)a^2(\tau),
\end{equation}
whose solution is given by
\begin{equation}\label{39}
a^6(\tau)={24c_1c_2 e^{\pm\sqrt{6c_2\over c_1}(C-\tau)}\over
\left(12c_1c_2^2+e^{\pm\sqrt{6c_2\over c_1}(C-\tau)}\right)^2},
\end{equation}
Where $C$ is a constant.  Using $h={1\over
6a^6(\tau)}{da^6(\tau)\over d\tau}$ one obtains
\begin{equation}\label{40}
h=\left(\pm\sqrt{c_2\over 6c_1}\right)\left({-12c_1c_2^2+e^{\pm\sqrt{6c_2\over c_1}(C-\tau)}\over 12c_1c_2^2+e^{\pm\sqrt{6c_2\over c_1}(C-\tau)}}\right).
\end{equation}
As $h$ is a decreasing function of $\tau$, we expect that $h$
becomes negative after some time, dubbed as turnaround time (see
fig(1) and fig(2)). This turnaround occurs at
\begin{equation}\label{41}
\tau=C\mp\sqrt{c_1\over 6c_2}\ln(12c_1c_2^2).
\end{equation}
As crossing the phantom divide line is not permitted, the Hubble
parameter continues to decrease and reach at
\begin{equation}\label{42}
h(\tau\to\infty)=-\sqrt{c_2\over 6c_1},
\end{equation}
asymptotically.

To study the acceleration of the universe, we consider
\begin{equation}\label{43}
S:={d{h}\over d\tau}+h^2,
\end{equation}
which has the same sign as $\ddot{a}$. $S=0$ occurs at the times
$\tau_1$ and $\tau_2$ specified by
\begin{eqnarray}\label{44}
\tau_1&=&C\mp\sqrt{c_1\over 6c_2}\ln(167.14c_1c_2^2)\nonumber \\
\tau_2&=&C\mp\sqrt{c_1\over 6c_2}\ln(0.859c_1c_2^2).
\end{eqnarray}
To elucidate the behavior of $S$, we compute  ${dS\over d\tau}$ at
these points. The result is
\begin{eqnarray}\label{45}
{dS\over d\tau}(\tau_1)&=&\mp 0.35354\left({c_2\over c_1}\right)^{3\over 2}\nonumber \\
{dS\over d\tau}(\tau_2)&=&\pm 0.35285\left({c_2\over
c_1}\right)^{3\over 2},
\end{eqnarray}
which shows that in this model the universe has a positive
acceleration (in the sense that $\ddot{a}>0$) for $\tau<
min.\{\tau_1,\tau_2\}$ and $\tau> max.\{\tau_1,\tau_2\}$. For
$min.\{\tau_1,\tau_2\}<\tau<max.\{\tau_1,\tau_2\}$ we have
$\ddot{a}<0$.

So far, as we have not fixed the values of $c_1$, $c_2$, and $C$,
our discussions and results were general.  To be more specific, we
must put some physical conditions on these parameters. To do so,
we consider the effective EoS parameter of dark sector,
$w_T={P_T\over \rho_T}$, where $P_T$ and $\rho_T$ are the
effective pressure and energy density attributed to modified
teleparallel gravity respectively \cite{lin}. We also consider the
relative densities defined by $\Omega_m={8\pi \rho_m\over 3H^2}$,
and $\Omega_T={8\pi \rho_T\over 3H^2}$. The relative dark matter
density can be written as
\begin{equation}\label{46}
\Omega_m={a^{-3}\over 6h^2}.
\end{equation}
Using $w=\Omega_T w_T$\cite{hao}, and $\Omega_m+\Omega_T\approx
1$, we obtain
\begin{equation}\label{47}
w_T={-4 {dh\over d\tau}-6h^2\over 6h^2-a^{-3}}.
\end{equation}
Equations (\ref{46}) and (\ref{47}) provide us a tool to determine
the parameters of the model. We take the present time to be at
$\tau=0$ and take \cite{wmap}
\begin{eqnarray}\label{48}
&&\Omega_m(\tau=0)\approx 0.25 \nonumber \\
&&w_T (\tau=0)\approx -1,
\end{eqnarray}
as initial conditions (i.e. conditions respected by the model in a
specific time, e.g. the present epoch). By solving (\ref{48}),
$c_1$ and $c_2$ are obtained in terms of $C$ as
\begin{eqnarray}
&&c_1 ={0.04572W(-1.5951\left|C\right|))^2\over C^2}\nonumber \\
&&c_2 = {0.7621W(-1.5951\left|C\right|)^4\over C^4},
\end{eqnarray}
or
\begin{eqnarray}
&&c_1 = {0.04572W(0.788\left|C\right|)^2\over
C^2}\nonumber \\
&&c_2 = {0.7621W(0.788\left|C\right|)^4\over C^4},
\end{eqnarray}
where $W$,  denotes Lambert$W$ function. Our choices for $C$ may
be restricted by adopting further conditions. One of these
conditions may be that the dark sector of the universe encounters
a phase transition from $w_T>-1$ to $w_T<-1$ in the present era,
implying
\begin{equation}
{dw_T\over d\tau}(\tau=0)<0.
\end{equation}
We must also take into account $h(\tau=0)>0$ and $S(\tau=0)>0$,
corresponding to the fact that the present expansion of the
universe has a positive acceleration.

\begin{figure}[H]
\centering\epsfig{file=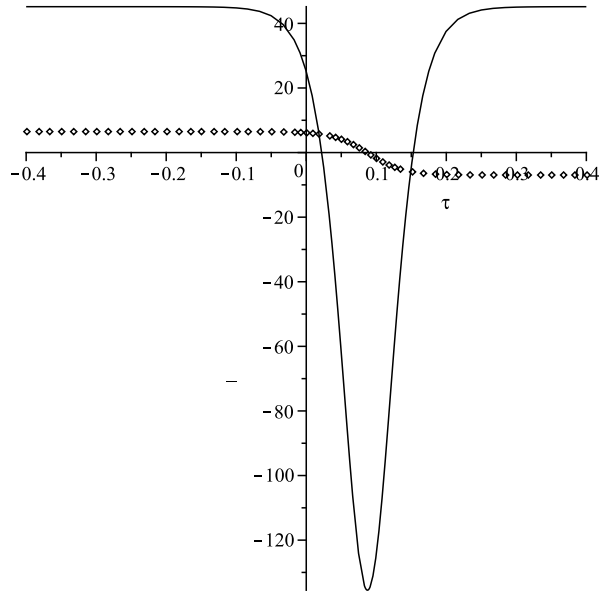,width=8cm} \caption{$S$ (line) and
$h$ (points) in terms of dimensionless time $\tau$, for $\{c_1 =
0.7435, c_2 = 201.5087, C=-0.23\}$ corresponding to solution with
negative sign in (\ref{39}).} \label{fig1}
\end{figure}

\begin{figure}[H]
\centering\epsfig{file=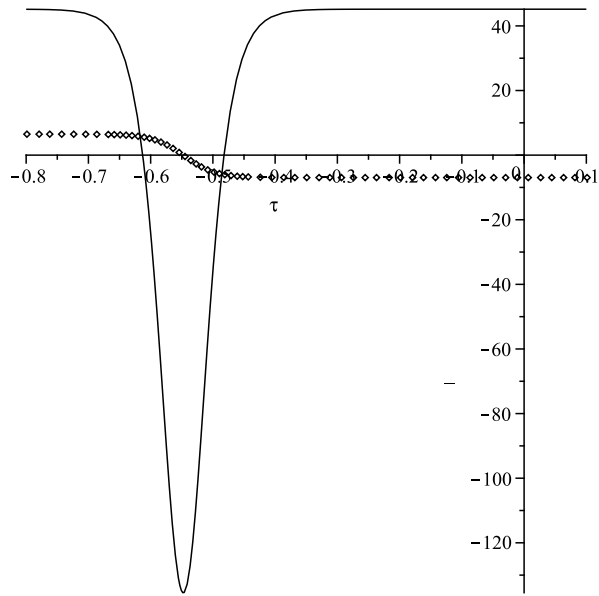,width=8cm} \caption{$S$ (line) and
$h$ (points) in terms of dimensionless time $\tau$, for $\{{c_1 =
0.7435, c_2 = 201.5087, C=-0.23}\}$ corresponding to solution with
positive sign in (\ref{39}). } \label{fig2}
\end{figure}

As an illustration of our results, in fig.(\ref{fig1})
(fig.(\ref{fig2})), $S$ and $h$ are depicted in terms of $\tau$,
for $\{c_1 = 0.7435, c_2 = 201.5087, C=-0.23\}$ and for solutions
with negative (positive) sign in (\ref{39}). Fig. (\ref{fig2}), as
is related to a contacting universe at $\tau=0$, does not
represent our actual universe.  In fig.(\ref{fig1}), we have
$\rho_{m0}\approx 0.0005H^2(\tau=0)$, where $H(\tau=0)={7\over
30000}Mpc^{-1}$ is the value of the Hubble parameter at the
present era \cite{wmap}. As we have not taken $a(0)=1$,
$\rho_{m0}$ must not be confused with the value of matter density
at the present time.

Note that, the qualitative behaviors of $h$ and $S$ in these
figures are consistent with was discussed after eq.(\ref{45}), and
do not depend on the specific values of the parameters. But as we
have shown, astrophysical data, by fixing the initial conditions,
put some quantitative constraints on these behaviors.

\begin{figure}[H]
\centering\epsfig{file=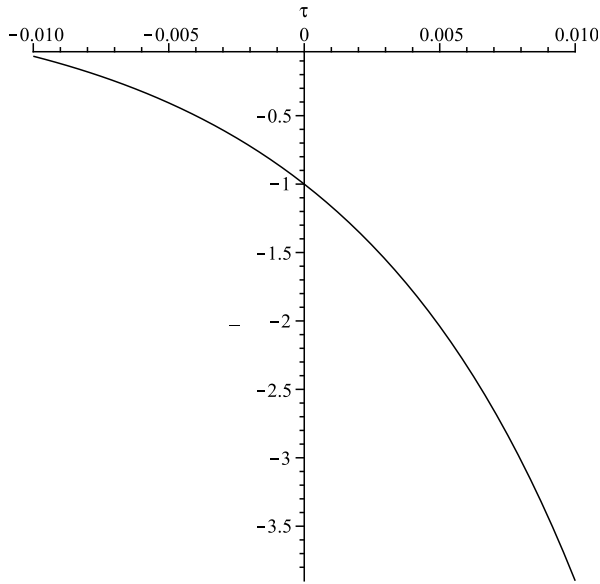,width=8cm} \caption{$w_T$ in terms
of dimensionless time $\tau$, for $\{{c_1 = 0.7435, c_2 =
201.5087, C=-0.23}\}$ corresponding to solution with negative sign
in (\ref{39}). } \label{fig3}
\end{figure}

The explicit form of crossing the line $w_T=-1$, in the present
era, is depicted in fig.(\ref{fig3}).
This figure shows that, by a suitable choice of the parameter $C$,
the dark sector can exhibit a transition from $w_T>-1$ to $w_T<-1$
in the present era.

\section{Conclusion}
The modified teleparallel gravity ($f(T)$ model of gravity) is a
framework to study the present accelerated expansion of the
universe. To determine the form of $f(T)$, one can use the Noether
symmetry. This method was vastly used in the literature to specify
the form of modifications in modify theories of gravity as well as
to determine the form of the scalar field potential in dark energy
models. In \cite{hao}, where the coefficients of vector fields are
taken to be functions of coordinates, only one form for $f(T)$ is
deduced: $f(T)=\mu T^n$, where $\mu$ and $n$ are two real numbers.
The scale factor is $a(t)\propto t{2n\over3}$ and consequently the
Hubble parameter is obtained as $H(t)={2n\over 3t}$. Therefore the
EoS parameter of the universe is a constant $w={1\over n}-1$, so
$n>{3\over 2}$ results in an eternal acceleration for the
expanding universe.

In this paper, this approach was extended to the generalized
Noether symmetry where the coefficients of generalized vector
fields contain the terms linear in velocities (time derivative of
configuration space coordinates). In our study, we assumed that
the matter component of the universe is dominated by cold dark
matter. Beside the solution of \cite{hao}, we obtained new
explicit forms for $f(T)$ (see (\ref{30})). The novel aspects of
these solutions are that the EoS parameter of the universe is no
more a constant, and besides a turn around, the universe
encounters successive acceleration deceleration phases in the era
where the matter component is dominated by cold dark matter.

In our model the universe cannot cross the phantom divide line.
Indeed, as was explained, we have always $\dot{H}<0$, whence
$w>-1$. Despite this, the EoS parameter of the dark sector, namely
$w_T$,  may be lower than or equal to $-1$, and $w_T>-1$ to
$w_T<-1$ transition, as illustrated in fig.(\ref{fig3}), may occur
in our present era. This characteristic is absent in some previous
frameworks where an exponential form, or other forms inspired by
viable $f(R)$ gravities, for $f(T)$ were proposed \cite{cross}. As
a summary, on the base of fundamental symmetry theories,
generalizing the Noether symmetry provides us a mean to obtain new
forms for $f(T)$, which may be physically more viable.

In our study the coefficients in vector fields were linear in the
velocities (see eq.(\ref{22})). By adopting more generalized form
of vector fields, it may be quite possible to obtain new $f(T)$
models which may be more consistent with astrophysical data.
However, due to computational hurdles, finding analytical
solutions in these cases is very complicated.


\begin{thebibliography}{99}

\bibitem{tel} A. Unzicker and T. Case, arXiv:physics/0503046 [physics.hist-ph];
C. Pellegrini and J. Plebanski,  K. Dan. Vidensk. Selsk. Mat. Fys.
Skr. 2, No. 2 (1962);  K. Hayashi and T. Nakano, Prog. Theor.
Phys. 38 (1967) 491 ; V. C. de Andrade, L. C. T. Guillen, and J.
G. Pereira, arXiv:gr-qc/0011087.
\bibitem{connect} R.Weitzenb\"{o}ck, Invarianten Theorie, (Nordhoff, Groningen,
1923).
\bibitem{acc}S. Perlmutter et al, Nature (London) 391 (1998) 51;
E. J. Copeland, M. Sami, and S. Tsujikawa, Int. J. Mod. Phys. D 15
(2006) 1753.
\bibitem{modif}A. D. Felice and S. Tsujikawa, Living Rev.
Rel. 13 (2010) 3 ; T. P. Sotiriou and V. Faraoni, Rev. Mod. Phys.
82 (2010) 451; S. Nojiri and S. D. Odintsov, Int. J. Geom. Meth.
Mod. Phys. 4 (2007) 115; S. Nojiri and S.D. Odintsov, Phys. Rept.
505, 59 (2011); H. M. Sadjadi, Phys. Rev. D77 (2008) 103501; K.
Bamba, S. Capozziello, S. Nojiri, and S. D. Odintsov, Astrophys.
Space Sci. 342 (2012) 155 ; S. Capozziello and M. D. Laurentis,
Phys. Rept. 509 (2011) 167.
\bibitem{teled}
E. V. Linder, Phys. Rev. D 81 (2010) 127301 ; R. Ferraro and F.
Fiorini, Phys. Rev. D 75 (2007) 084031 ; B. Li, T. P. Sotiriou,
and J. D. Barrow, Phys. Rev. D 83 (2011) 104017; M. Li, R. X.
Miao, and Y. G. Miao,  JHEP 1107 (2011) 108; P. Wu and H. Yu,
Phys. Lett. B 693 (2010) 415; H. Wei, H. Y. Qi, and X. P. Ma, Eur.
Phys. J. C 72 (2012) 2117; H. Wei, X. P. Ma, and H. Y. Qi, Phys.
Lett. B 703, 74 (2011 ); C. Q. Geng, C. C. Lee, E. N. Saridakis,
and Y. P. Wu, Phys.  Lett. B 704 (2011) 384 ; M. E. Rodrigues, M.
J. S. Houndjo, D. Saez-Gomez, and F. Rahaman, arXiv:1209.4859
[gr-qc]; M. J. S. Houndjo, D. Momeni, and R. Myrzakulov,
arXiv:1206.3938 [physics.gen-ph]; K. Bamba, M. Jamil, D. Momeni,
and R. Myrzakulov, arXiv:1202.6114 [physics.gen-ph]; K. Karami,
and A. Abdolmaleki, JCAP 04 (2012) 007; A. Behboodi, S. Akhshabi,
and K. Nozari,  arXiv:1205.4570 [gr-qc]; K. Karami, and A.
Abdolmaleki, arXiv:1202.2278 [physics.gen-ph]; M. H. Daouda, M. E.
Rodrigues, and M. J. S. Houndjo, Eur. Phys. J. C 72 (2012) 1893;
M. Sharif, and S. Rani, Phys. Scr. 84 (2011) 055005; A. Banijamali
and B. Fazlpour, arXiv:1206.3580 [physics.gen-ph]; S. Capozziello,
V. F. Cardone, H. Farajollahi, and A. Ravanpak, Phys. Rev. D 84
(2011) 043527; X. Meng, Y. Wang, arXiv:1107.0629 [astro-ph.CO]; K.
Bamba, R. Myrzakulov, S.  Nojiri, and S. D. Odintsov, Phys. Rev. D
85 (2012) 104036.
\bibitem{lin}E. V. Linder, Phys. Rev. D 81 (2010) 127301.
\bibitem{mir}R. Myrzakulov,  Eur. Phys. J. C 71 (2011) 1752.
\bibitem{hao}H. Wei, X. J. Guo, and L. F. Wang, Phys. Lett. B 707 (2012) 298.
\bibitem{quinmod}R. de Rittis, G. Marmo, G. Platania, C. Rubano, P. Sudellaro, and S. Stornaiolo,
Phys. Rev. D 42 (1990) 1091; B. Modak, S. Kamilya, and S. Biswas,
Gen. Rel. and Grav. 32 (2000) 1615; Y. Zhang, Y. G. Gong, and Z.
H. Zhu, Phys. Lett. B 688 (2010) 13; S. Basilakos, M. Tsamparlis,
and A. Paliathanasis, Phys.Rev.D 83 (2011) 103512 ; S.
Capozziello, E. Piedipalumbo, C. Rubano, and P. Scudellaro, Phys.
Rev. D 80 (2009) 104030; S. Basilakos, M. Tsamparlis, and A.
Paliathanasis, Phys. Rev. D 83 (2011) 103512; S. Capozziello and
A. D. Felice, JCAP 0808 (2008) 016 ; B. Vakili, Phys. Lett. B 664
(2008) 16 ; M. Jamil, F. M. Mahomed, and D. Momeni, Phys. Lett. B
702 (2011) 315 ; S. Capozziello, E. Piedipalumbo, C. Rubano, and
P. Scudellaro, Phys. Rev. D 80 (2009) 104030 ; S. Capozziello, M.
D. Laurentis, and S. D. Odintsov, Eur. Phys. J. C 72 (2012) 2068 ;
I. Hussain, M. Jamil, and F. M. Mahomed, Astrophys. Space Sci.
337, (2012) 373; M. Jamil, S. Ali, D. Momeni, and R. Myrzakulov
Eur. Phys. J. C 72 (2012) 1998; M. Jamil, D. Momeni, and R.
Myrzakulov Eur. Phys. J. C  72 (2012) 2137; K. Atazadeh, and F.
Darabi, Eur. Phys. J. C 72 (2012) 2016; B. Vakili and F. Khazaie,
Class. Quantum Grav. 29 (2012) 035015.
\bibitem{GN} L. Fatibene, M. Ferraris, M. Francavigilia, and R. G. McLenagaghan, Jour. Mth. Phys. 43(2002)3147.
\bibitem{kep} L. Fatibene, M. Francaviglia, and S. Mercadante, arXiv:1001.2886 [gr-qc].
\bibitem{wmap}E. Komatsu et al., Astrophys. J. Suppl. 192, 18 (2011), arXiv:1001.4538
[astro-ph.CO].
\bibitem{cross}K. Bamba., C. Geng,  C. Lee, and L. Luo, JCAP 01,
(2011) 021, arXiv:1011.0508 [astro-ph.CO]; K Bamba, C. Geng, and
C. Lee, arXiv:1008.4036 [astro-ph.CO].


\end{thebibliography}
\end{document}